# Robust Flat-Magnetoresistivity in D0₃-Fe₃Ga Driven by Chiral Anomaly

Ruoqi Wang[1]*, Xinyang Li[1]*, Bo Zhao[2,3]*, Haofu Wen[1]*, Xin Gu[1], Shijun Yuan[1], Langsheng Ling[4], Chuanying Xi[4], Ze Wang[4], Kunquan Hong[1], Liang Ma[1,5]✉, Ke Xia[1]✉, Taishi Chen[1]✉, Jinlan Wang[1,5]

1. *Key Laboratory of Quantum Materials and Devices of Ministry of Education, School of Physics, Southeast University, Nanjing 21189, China*

2. *Shanghai Advanced Research Institute, Chinese Academy of Sciences, Shanghai 201204, China*

3. *Henan Key Laboratory of Imaging and Intelligent Processing, Information Engineering University, Zhengzhou 450001, China*

4. *Anhui Key Laboratory of Low-Energy Quantum Materials and Devices, High Magnetic Field Laboratory, HFIPS, Chinese Academy of Sciences, Hefei, Anhui 230031, China*

5. *Suzhou Laboratory, Suzhou 215004, China*

*\* Ruoqi Wang, Xinyang Li, Bo Zhao and Haofu Wen contributed equally.*
*✉Corresponding author. E-mail:* chentaishi@seu.edu.cn, *kexia@seu.edu.cn, liang.ma@seu.edu.cn*

## Abstract

Topologically non-trivial nodes emerging from flat-band crossings not only enhance unconventional topological responses but also play a fundamental role in exploring correlation-driven topological physics. Here, we report the exceptionally robust chiral-anomaly-dominated transport in $D0_3$-$Fe_3Ga$. Firstly, we observe a perfect combination of positive/negative magnetoresistance, ideal planar longitudinal magnetoresistance (PLMR), and planar Hall effect (PHE). Secondly, ultra-low-temperature resistivity exhibits pronounced non-Fermi-liquid (NFL) behavior, accompanied by the emergence of giant intrinsic anomalous Hall conductivity (AHC), in excellent agreement with our DFT calculations, which confirm the existence of tilted Weyl points arising from crossings of nearly three-dimensional (3D) flat-bands. Most remarkably, we detect an exceptionally robust flat-magnetoresistance (flat-MR) that persists without decay up to 33 T. This distinctive set of phenomena provides strong evidence that the Fermi level intersects the flattened Weyl crossings, offering unambiguous confirmation of a long-sought hallmark of ideal topological semimetals. Unlike previous systems, $D0_3$-$Fe_3Ga$ exhibits the pristine characteristics of a topological flat-band semimetal. Our findings also highlight a promising magnetic platform for quantum device innovations.

## Introduction

When two nearly flat bands come into proximity and cross in $k$-space, the protection provided by the corresponding crystal symmetries can lead to topologically non-trivial nodal-crossing bands, giving rise to a wide variety of chiral fermions. Moreover, these chiral fermions not only possess well-defined Chern topology but may also bring the characteristic electronic correlations associated with flat bands, rendering them distinct from those in conventional single-electron systems[1, 2]. In particular, when such flat-band crossings lie in proximity to the Fermi level, they can contribute an exceptionally high density of states (DOS), thereby dominating the topological transport behaviors in topological materials. Compared with great achievements in 2-dimensional (2D) flat-band systems[2, 3, 4, 5, 6], the progress in the study of 3D nodal flat-band systems has lagged. 3D nodal flat-band materials provide an exceptional platform for exploring exotic electronic phenomena[7]. Firstly, the correlation among nodal flat-band electrons may challenge the conventional single-electron model that defines the topology of condensed materials and unique electromagnetic phenomena, and particularly exhibits NFL behavior[8, 9, 10, 11]. Secondly, magnetic 3D nodal flat-band electrons are characterized by low mobility and high DOS, significantly amplifying the intrinsic anomalous Hall effect (AHE)[7]. Finally, the presence of abundant chiral electrons in such systems naturally leads to pronounced positive and negative MR[12, 13, 14, 15, 16, 17, 18, 19], as well as perfect PHE and PLMR related to chiral anomaly [20, 21, 22, 23, 24].

However, discovering such an ideal 3D nodal flat-band topological semimetal presents significant challenges. While extensive theoretical work has predicted 3D nodal flat-band topological semimetals, such as Kagome systems[25, 26] and pyrochlore lattices[7, 8], the challenges associated with the realization of these materials remain exceedingly difficult. Firstly, the crystal quality of such materials must be exceptionally high to minimize shifts in the Fermi level caused by defects[27]. Additionally, the electron correlation in these materials may result in the breakdown of the topological description derived from single-particle models[8, 28]. Furthermore, the Fermi level must reside in close proximity to the nodal-band crossings. These stringent requirements present insurmountable obstacles in observing the unique magnetoelectric response associated with the 3D nodal flat-band. Even within 3D Dirac/Weyl semimetals, these challenges also persist[23, 29, 30, 31].

Recently, a nearly 3D flat-band in D0$_3$-phase Fe$_3$Ga has attracted great attention, which is estimated to lie only 74 meV below the Fermi level by matching the giant room-temperature anomalous Nernst effect with the DFT results[10]. Due to the extreme difficulty in synthesizing high-quality bulk single crystals[32], research on the topological properties of these 3D flat-band remains largely unexplored to date. Here, we report our breakthrough on this issue. Comprehensive MR measurements reveal that it exhibits significant differences from currently benchmark topological semimetals[29, 33, 34]. Crucially, we discovered a robust flat-MR in the cryogenic temperatures under a high-magnetic field as high as 33 T, without any indication of decay. Furthermore, the giant AHC of more than 1400 S/cm is also obtained, which is 200 S/cm higher than the maximum value predicted in Ref. [10]. Importantly, our DFT calculations confirm that the enhancement of the AHC is directly linked to the formation of tilted Weyl cones close to the location of the original nodal web. Finally, the perfect PLMR and PHE induced by the chiral quasiparticles and their transition to an anisotropic MR (AMR) scenario are also successfully observed[35]. All of these unique experimental achievements, along with the NFL behavior of the low-temperature resistivity, collectively affirm that the D0$_3$-phase Fe$_3$Ga is a prototypical 3D nodal flat-band ferromagnet. Our findings may provide a promising magnetic material for quantum device innovations, such as the 3D anomalous Hall effect device[36].

**Basic characterizations and ultra-low temperature RT in D0$_3$-phase Fe$_3$Ga**

Figure 1a shows the flat-band structure fostering Weyl points in D0$_3$–Fe$_3$Ga. It is important to emphasize that the lattice constant used in our calculations was obtained from our own TEM measurements and analysis. Before including SOC, our DFT-calculated band structure is in good agreement with previously reported results (Supplementary Information Fig. S14)[10]. In particular, we identify at the $L$ point a nodal web formed by two nearly flat-band crossings, as indicated in the black dashed box. Upon inclusion of SOC, the nodal web is destroyed, opening a gap of about 10 meV. This gap is confirmed to be topologically trivial, as no traces of surface or edge states are present within it. Crucially, and in sharp contrast to previous work[10], we find the emergence of Weyl points along the $L$–$W$ direction. The corresponding Weyl cones span a narrow energy window of about 10 meV and exhibit a pronounced tilt. They directly connect to the flat-band segments originating from the nodal web. We find that such flat-band-derived Weyl points, when lying in proximity to the Fermi level, can generate a giant AHC up to 2200 S/cm, contribute a substantial density of chiral fermions, and produce an even more pronounced chiral-anomaly-induced magnetoresistance (Supplementary Information Fig. S11, S12, S13, and S15)).

To experimentally probe the magnetotransport signatures arising from these flat-band-derived Weyl points, we worked on a high-quality Fe$_3$Ga bulk single crystal obtained by the chemical vapor transport method (CVT, see Methods). Fig. 1b shows the Fe$_3$Ga bulk single crystals. The TEM image shown in Figure 1c displays the (111) peaks, which verify that these crystals belong to the D0$_3$ phase rather than the lower-symmetry B$_2$ and A$_2$ phases (Supplementary Information Fig. S1)[10], and yield a lattice constant of c=6.13 Å. Notably, the TEM signals are very sharp and do not show satellite spots, indicating higher crystal quality surpassing crystals in *Ref*[10]. The improved crystal quality is further confirmed by the residual resistivity ratio (RRR) of 5, as shown in Fig. 1f, which is nearly three times higher than that reported in previous work[10]. For temperatures between 4 K and 60 mK, the resistivity was analyzed using representative $R^2$ values from a power-law fitting to $\rho(T) = \rho_0 + AT^n$ as a function of $n$, as shown in the inset of Fig. 1f. The maximum $R^2$ value (black dots) indicates optimal fitting, yielding n$_{opt}$=1.53. At the same time, the parameters $A$ and $\rho_0$ were obtained as $4.42\times10^{-4}$ μΩ cm/K$^2$ and 10.82 μΩ cm, respectively. This deviation from the Fermi-liquid value ($n = 2$) strongly suggests the presence of electron correlation effects in D0$_3$-Fe$_3$Ga[9], as also seen in Supplementary Information Fig. S9. Furthermore, the M–T curve shown in Fig. 1e indicates that the Curie temperature is higher than 800 K. The Cuire temperature exceeds the 720 K reported in *Ref*[10], and is consistent with the previously extrapolated value of ~1000 K[37], further demonstrating the significant material advantages of high-quality D0$_3$-Fe$_3$Ga for future high-performance quantum devices. The *M vs. H* curves measured at 5 K and 300 K reach saturation at 0.4 T and do not display any anomaly in the saturated state (Fig. 1d). These results establish the high quality of our D0$_3$-phase Fe$_3$Ga crystals, enabling a reliable investigation of their flat-band–driven topological transport. As the flat-band Weyl points produce strong Berry curvature, we first focus on the AHC.

**Giant anomalous Hall conductivity in D0$_3$-Fe$_3$Ga**

Fig. 2a shows the Hall resistivity measured at different temperatures. The anomalous Hall resistivity undergoes a twentyfold increase from 0.15 mΩ cm at 2 K to 3.2 mΩ cm at 300 K. Strikingly, the AHC obtained using the formula: $-\sigma_{yx} = \frac{\rho_{yx}}{\rho_{xx}^2+\rho_{yx}^2}$, not only reaches up to 1400 S/cm (Fig. 2b), two times higher than the value in *Ref*[10], but also 200 S/cm exceeds by the maximum value predicted by the DFT method in *Ref*[10]. Besides, this giant AHC persists in the whole temperature range without obvious decline, even when extrinsic magnetic scattering becomes dominant above 50 K (Supplementary Information Fig. S5). Moreover, both AHE and AHC show a clear slope transition. In nodal semimetals, this is often accompanied by the unique longitudinal magnetoresistivity. In Fig. 2c, we show the MR and Hall resistivity up to 33 teslas at 1.8 K. Below 12 T, the MR exhibits a quadratic field dependence, transitioning to a linear regime that extends up to 33 T. Such a crossover from parabolic to linear MR closely resembles the magnetotransport signatures well-established in benchmark topological semimetals[15, 33, 34, 38, 39], ruling out the positive MR induced by the magnetic scattering in the low temperature regime[40]. To determine the mobility and charge carrier density, we used the Drude two-carrier model to fit the Hall resistivity $\rho_{yx}$ at 1.8 K, such that $\rho_{yx} = -\frac{1}{e}\frac{(n_e\mu_e^2-n_h\mu_h^2)B+\mu_e^2\mu_h^2(n_e-n_h)B^3}{(n_e\mu_e+n_h\mu_h)^2+\mu_e^2\mu_h^2(n_e-n_h)^2B^2}$, where $n_e$ and $\mu_e$ refer to electron carrier density and mobility, respectively. The fit is shown in Fig. 2d, yielding majority electrons:

$n_e = 4.25 \times 10^{23} \ cm^{-3}$, $\mu_e = 10 \ cm^2 V^{-1} s^{-1}$; minority holes: $n_h = 2.27 \times 10^{20} \ cm^{-3}$, $\mu_h = 4300 \ cm^2 V^{-1} s^{-1}$.

In semimetals, ultrahigh carrier mobility of $> 1000 \ cm^2 V^{-1} s^{-1}$ often induces high positive MR for $\hat{\mathbf{B}} \perp \hat{\mathbf{I}}$, arising from charge compensation, as well as negative MR under $\hat{\mathbf{B}} \parallel \hat{\mathbf{I}}$ due to the extrinsic current-jetting effect[29, 41]. In contrast, the moderate carrier mobility and high carrier density in Fe₃Ga can effectively suppress this extrinsic effect, which is also confirmed by our squeezing test (Supplementary Information Fig. S8)[29, 34]. The exceptionally large AHC, the slope transition observed in the Hall effect, and the distinctive MR under high steady magnetic fields together provide compelling evidence that the Fermi level nearly intersects Weyl points (Supplementary Information Fig. S12). Therefore, the present D0₃-Fe₃Ga crystal offers an ideal platform to study the MR response specific to the chiral anomaly, which is discussed in the next section.

**The robust flat-MR in D0₃-Fe₃Ga driven by chiral anomaly**

Fig. 3a shows the MR curves at 1.8 K, with the magnetic field changing from [010] to the [100] direction. Compared with previous achievements in nodal semimetals[30], the D0₃ phase Fe₃Ga single crystal shows more intriguing MR related to the topological flat band nodals. Firstly, under $\hat{\mathbf{B}} \perp \hat{\mathbf{I}}$, the MR increases as $B^2$ up to 12 T, followed by a linear increase to 33 T without any indication of decrease or saturation. Secondly, for $\hat{\mathbf{B}} \parallel \hat{\mathbf{I}}$ the MR exhibits a near $\mathbf{B^2}$ dependence in the high-field regime, showing no signs of either upturn or saturation. This behavior underscores the robustness of the chiral-anomaly-induced quasiparticle pumping mechanism[14, 17]. Thirdly, we observed a striking flat-MR at angles near $45^0$ and $135^0$, which is very rarely reported even in the benchmark nodal semimetals, including Na₃Bi[34, 42], TaAs[15, 43, 44, 45], Co₃Sn₂S₂[46, 47], Mn₃Sn[48], and Co₂MnGa[11, 49, 50]. We further found that the demagnetization effect can significantly influence the angles at which flat-MR occurs (see SI Section S6 and Fig. S10). To observe such unique MR, three conditions must be met simultaneously in one sample: 1) the Fermi surface must lie in very close proximity to the Weyl points; 2) extrinsic effects such as magnetic scattering, current-jetting, AMR, etc., must be completely suppressed; and 3) the positive MR for $\hat{\mathbf{B}} \perp \hat{\mathbf{I}}$, must be comparable with the negative MR under $\hat{\mathbf{B}} \parallel \hat{\mathbf{I}}$.

To further exclude the possibility of a current-jetting artifact, we conducted the standard squeezing test[29]. As expected, under $\widehat{\mathbf{B}} \parallel \widehat{\mathbf{I}}$ both the edge and the central contacts of the crystal demonstrate identical negative MR behavior (Supplementary Information Fig. S8). This observation unambiguously corroborates that the negative MR in the high-quality $D0_3$-phase $Fe_3Ga$ single crystal originates from the chiral anomaly rather than from extrinsic measurement artifacts.

The results shown in Fig.3a strongly suggest $Fe_3Ga$ of $D0_3$-phase is an ideal ferromagnetic topological semimetal. The negative MR persists across a wide range of angles, in sharp contrast to the previous observations of the benchmark magnetic nodal semimetals. We also exclude the weak/anti-localization (WL/WAL) effect happening in $Fe_3Ga$[51]. It is well known that WL- and WAL-induced MRs in nodal semimetals exhibit turning points at a moderate magnetic field under $\widehat{\mathbf{B}} \parallel \widehat{\mathbf{I}}$[24, 51], which contrasts strongly with our observations. Therefore, our observation presents an unprecedented opportunity for studying two key effects related to the chiral anomaly, namely, the PHE and PLMR.

Fig. 3b shows the PHE and PLMR induced by the 3D nodal flat-band under $\mathbf{B} = 9T$ at 2 K and 300 K, respectively. Here, the $\theta$ is defined as the angle between the magnetic field and Hall voltage in the Hall-current plane. The background-subtracted PLMR ($\Delta\rho^{PLMR}$) is defined as: $\Delta\rho^{PLMR}(\theta) = \rho_{xx}^{PLMR}|_{\mathbf{B}=9T}(\theta) - \rho_{xx}^{PLMR}|_{\mathbf{B}=0T}$. $\rho_{xx}^{PLMR}$ refers to the longitudinal resistivity measured when the magnetic field $\widehat{\mathbf{B}}$ is rotated within the Hall-current plane, under fixed magnetic field and temperature. Such distinctive behaviors are reminiscent of previous works on Weyl semimetals[20, 22], where only the chiral-anomaly–induced PHE and PLMR are considered. The $\rho_{chiral}^{PHE}$ and $\rho_{chiral}^{PLMR}$ are expressed as[20, 22]:

$$\rho_{chiral}^{PHE} = -\Delta\rho_\chi \sin\left(\frac{\pi}{2} - \theta\right)\cos\left(\frac{\pi}{2} - \theta\right) \qquad (1)$$

$$\rho_{chiral}^{PLMR} = \rho_0 - \Delta\rho_\chi \cos^2\left(\frac{\pi}{2} - \theta\right) \qquad (2)$$

Where $\rho_0$ is consistently defined as the longitudinal resistivity under zero magnetic field (B=0), and the $\Delta\rho_\chi$ is defined as $\Delta\rho_\chi = \rho_{xx}(0T) - \rho_{xx}(9T)|_{\widehat{\mathbf{B}}\parallel\widehat{\mathbf{I}}}$. It should be emphasized that, directly using the two formulas to fit the experimental data can lead to misleading conclusions, as the two formulas do not account for the positive MR induced by the various nodal crossing bands[11, 21, 23, 46, 48]. Here, we use the modified formulas to fit the $\rho^{PHE}(\theta)$ and $\Delta\rho^{PLMR}(\theta)$:

$$\rho_{modified}^{PHE}(\theta) = \Delta\rho_\chi \sin(2\theta) \qquad (3)$$

$$\Delta\rho_{modified}^{PLMR}(\theta) = \Delta\rho_\chi \sin\left(2\theta + \frac{\pi}{2}\right) \qquad (4)$$

and yield excellent fits to the experimental data, as shown in Fig.3b. The key distinction between the four formulas lies in that the amplitude in (3) and (4) is exactly twice as large as that in (1) and (2), due to the equal amplitude of the positive MR induced by the Weyl points. The derivation from equations (1) and (2) to equations (3) and (4) is provided in the Supplementary Information S4.

With temperature increasing, particularly beyond 50 K, the $\Delta\rho^{PLMR}(\theta)$ gradually diverges from the formula (4). The $\rho^{PHE}(\theta)$ still persists up to a very high temperature, even at 300 K. For 300 K，the $\Delta\rho^{PLMR}(\theta)$ shift forward exactly by a phase of $\pi/2$, fully evolving into the AMR regime. This is the ~~first~~ clear example that highlights the essential differences between the nodal crossing bands and the AMR-induced PHE and PLMR in 3$d$ semimetals. In so-far reported magnetic nodal semimetals, the 3$d$ transition elements, such as iron, cobalt, and manganese, play crucial roles in fostering topological states[52, 53, 54], but they also produce conventional MRs, particularly negative MR and AMR, which originate from the magnetic scattering and magnetically-created current effect[35, 40], respectively. Therefore, identifying the chiral anomaly that induces negative MR, PHE, and PLMR in magnetic nodal semimetals requires extreme caution. In Fe₃Ga, the magnetic scattering also becomes significant as the temperature increases over 50 K. These transitions can also be confirmed by checking the MR in Fig. 3c and 3d.

In summary, these observations in both PHE and PLMR again strongly support the evidence for the topological non-triviality of the flattened Weyl points in the D0₃-phase of Fe₃Ga.

**Competition between chiral anomaly and anisotropic magnetoresistance**

As shown above, the PLMR serves as a more reliable way in confirming chiral anomaly than PHE, particularly for magnetic nodal semimetals including 3$d$ transition elements. In Fig. 4a, we show the PLMR obtained at different temperatures. Using the same experimental configuration as in Figure 3b. Above 50 K, the initial PLMR induced by the flattened Weyl points gradually transitions into the version of AMR. To establish standard criteria for verifying the presence of nodal bands in topological semimetals via PLMR and PHE measurements, we investigate the phase difference between PLMR and PHE. The results are shown in Fig. 4b. Here, the phase shift is defined: $\Delta\Phi = \Phi^{PLMR} - \Phi^{PHE}$, where the $\Phi^{PHE}$ and $\Phi^{PLMR}$ are included below:

$$\rho^{PHE}(\theta, B_0) = \Delta\rho_\chi^{PHE} \sin2(\theta + \Phi^{PHE})) \qquad (5)$$

$$\rho^{PLMR}(\theta, B_0) - \rho(B_0) = \Delta\rho_\chi^{PLMR} \sin2(\theta + \Phi^{PLMR}) \qquad (6),$$

where $\Delta\rho_\chi^{PHE}$ and $\Delta\rho_\chi^{PLMR}$ refer to the amplitudes. $\mathbf{B_0}$ denotes the applied magnetic field associated with the dominant magnetoresistance mechanism in the transport response. Below 50 K, $\mathbf{B_0} = 0$, while for high temperatures, $\mathbf{B_0}$ corresponds to the applied magnetic field used in AMR measurements. When the chiral anomaly dominates the PHE and PLMR, the $\mathbf{B_0}$ equals 0 T, which leads to $\Phi^{PHE} = 0$, $\Phi^{PLMR} = \pi/4$, and yielding $\Delta\Phi = \pi/4$. When the AMR becomes dominated, $B_0$ refers to the applied magnetic field, $\Phi^{PHE} = 0$ and doesn't change, while the $\Phi^{PLMR}$ changes to $\pi/2$, giving rise to $\Delta\Phi = \pi/2$. Besides, the amplitude ratio doesn't change in the entire temperature range. The comparison of the PHE and PLMR induced by chiral anomaly and AMR is also discussed in Supplementary Information Fig. S6 and S7. Thus, the phase shift between the PHE and PLMR establishes a standard criterion in identifying the magnetotransport unique to the chiral anomaly in nodal semimetals.

## Summary and Discussions

In summary, we have successfully grown high-quality bulk single crystals of the high-temperature ferromagnetic Weyl semimetal D0₃–Fe₃Ga. Comprehensive magnetotransport measurements reveal that its giant AHC originates from flattened Weyl crossings situated in the position of Fermi energy, accompanied by comparable-magnitude positive and negative MR, PHE, and ideal PLMR at low temperatures. Crucially, the strong evidence for the chiral-fermion origin of the negative MR has been confirmed through a squeezing test. Most notably, we observed, to date, the most robust Weyl-fermion-induced flat MR under high magnetic fields up to 33 T, without any indication of decay. These clear advantages compared with previously reported topological semimetals are summarized in the Table provided in the Supplementary Information.

In understanding the chiral-fermion origin behind these unique magnetoresistance phenomena, we have made significant progress through DFT calculations based on our experimentally determined lattice constant: 1) We confirm that, without SOC, a nodal web is observed, whereas the inclusion of SOC leads to the emergence of tilted Weyl points at the Fermi level; 2) Importantly, these Weyl points arise from the crossing of two nearly flat bands, which further endows the Weyl points in $D0_3$–$Fe_3Ga$ with pronounced flat-band characteristics, crucial for transport dominated by chiral fermions.3) In addition, using a Berry-curvature-based approach, we obtain the highest intrinsic anomalous Hall conductivity reported to date, reaching 2200 S/cm, which exceeds the previously calculated value in *Ref*[10] by 1000 S/cm. Furthermore, this system exhibits the largest reported transverse thermoelectric conductivity factor to date, exceeding 500 A/(m·K)—over 100 times greater than the highest values previously documented[10, 11, 55, 56]. This distinctive Weyl-point configuration in $D0_3$–$Fe_3Ga$ underlies the experimentally observed giant AHC, comparable positive and negative MR, pronounced PHE, ideal PLMR at low temperatures, and the most robust flat MR in the cryogenic regime.

By comparing our calculated results with those of previous work[10], we find that the emergence of Weyl points in $D0_3$–$Fe_3Ga$ exhibits a pronounced dependence on the lattice constant. This strong sensitivity to lattice parameters is reminiscent of the type-II Weyl semimetal $MoTe_2$, whose extreme dependence on lattice constants originates from its proximity to a topological quantum critical point where the band crossings forming the Weyl points have vanishingly small dispersion along a specific momentum direction, rendering them intrinsically unstable to even infinitesimal lattice deformations[57]. The similarity arises because, in $D0_3$–$Fe_3Ga$, the Weyl points are likewise generated by the crossing of two nearly flat conduction and valence bands, rendering them highly sensitive to changes in the lattice constant. This property provides an ideal platform for exploring stress-induced topological phase transitions in topological materials.

## Methods

$D0_3$-$Fe_3Ga$ single crystals were synthesized via a two-step process. In the first step, polycrystalline $Fe_3Ga$ powder was obtained through a solid-state reaction using high-purity iron and gallium in a 3:1 atomic ratio. The mixture was sealed and reacted at 1000 °C for one week. In the second step, $Fe_3Ga$ single crystals were grown by the CVT method in a long double-zone furnace, with iodine employed as the transport agent. Guided by the binary alloy phase diagram[58], the source and sink temperatures were set to 1000 °C and 900 °C, respectively. After three weeks of growth, shiny as-grown crystals displaying a metallic luster were obtained, with the largest specimens measuring $5mm \times 2mm \times 2mm$. From more than 50 grown crystals, the two highest-quality samples were selected based on their RRR, as determined from RT measurements.

Before compositional analysis, the as-grown single crystals were finely polished to remove any surface contamination. Quantitative determination of the Fe–Ga atomic ratio was carried out by scanning electron microscopy coupled with energy-dispersive X-ray spectroscopy (SEM/EDS), yielding a stoichiometry of 3.07:0.97 (Supplementary Information Fig. S2). The crystallographic orientation of each specimen was established using a single-crystal X-ray diffractometer (Rigaku Oxford XtaLAB Synergy-i, XtaLAB MMO07HFMR).

Because high-RRR specimens were scarce and $Fe_3Ga$ is mechanically hard, powder X-ray diffraction (XRD) measurements could not be performed. Instead, transmission electron microscopy (TEM) was employed to confirm the crystal structure of the two highest-RRR single crystals. The procedure involved two steps. First, focused ion beam (FIB) milling was used to cut a regular lamella from the center of the best sample (Supplementary Information Fig. S3). Next, a series of selected-area electron diffraction (SAED) patterns were collected from the prepared lamellae using a commercial TEM (Tecnai G2 T20, FEI, USA) along different crystallographic orientations. All diffraction patterns were acquired with an exposure time of 0.1 s after proper alignment, and high-resolution TEM (HRTEM) images were recorded with a 1 s accumulation time.

Bar-shaped specimens were prepared for all transport measurements and studied using commercial systems, including a Physical Property Measurement System (PPMS, Quantum Design) and Oxford superconducting magnets. Magnetization measurements were performed with the vibrating sample magnetometer option of a DynaCool-9 T system at the RIXS endstation, BL09UC, Shanghai Synchrotron Radiation Facility (SSRF, China). Longitudinal and Hall voltages were recorded in a six-probe configuration with silver-epoxy contacts, while high-field magnetoresistance (MR) at 1.8 K was measured using the WM5 facility at the High Magnetic Field Laboratory, Hefei, China. Electrical resistance from 4 K down to 60 mK was measured using a dilution-refrigerator (DR) module integrated with the PPMS, employing a standard four-probe configuration to minimize contact-resistance effects. Samples were mounted in the DR sample chamber, and the temperature was precisely controlled by the PPMS with calibration against internal thermometers to ensure accuracy. All low-temperature measurements were conducted under high-vacuum conditions to suppress thermal exchange with the environment, ensuring the fidelity and reproducibility of the data.

**Computational methods and details.**

All structural and electronic properties are calculated by using the Vienna ab initio simulation package based on spin-polarized density functional theory (DFT) [59]. The exchange correlation potential is described with the Perdew-Burke-Ernzerhof (PBE) of the generalized gradient approximation (GGA) [60]. The projected argument wave (PAW) [61] potential is used to describe the ion-electron potential. The energy cutoff of the plane wave is 500 $eV$. The Brillouin zone uses a convergent $11\times11\times11$ k-point mesh for structural relaxation and $15\times15\times15$ for self-consistent field calculations. Subsequently, the anomalous Hall conductivity was calculated via Wannier interpolation on a significantly denser $200\times200\times200$ k-point grid to ensure the convergence of the Berry curvature integration. The experimentally determined lattice constant was adopted for the calculations, with internal atomic positions relaxed until the force and energy converge to $0.001 eV$/Å and $10^{-5}$ $eV$/atom. The Hall conductance and topological characteristics are calculated by using the maximum local Wannier function implemented in the wannier90 software package [62, 63]. The calculated electronic band structure, the 3D nodal flat band, and the comparison between the experimental and calculated AHC, along with the examination of the associated correlation effects, are presented in the Supplementary Information Fig.S10-S13.

# Data Availability

Relevant data supporting the key findings of this study are available within the article and the Supplementary Information file. All raw data generated during the current study are available from the corresponding authors upon request.

# Acknowledgements

The authors thank Professor Peiwen Wu for the fruitful discussion. Funding: This work was supported by the National Key Research and Development Program of China, Grant # 2023YFA1406600 awarded to Taishi Chen. The National Natural Science Foundation of China, Grant # 12274068 awarded to Taishi Chen. The Scientific Research Innovation Capability Support Project for Young Faculty, Grant # 2242026K20003 awarded to Liang Ma. The National Natural Science Foundation of China , Grant # 92577115 awarded to Liang Ma.

The open research fund of Key Laboratory of Quantum Materials and Devices (Southeast University), Ministry of Education, the National Key Research and Development Program of China (2022YFA1503103), Natural Science Foundation of Jiangsu Province, Major Project (BK20222007), the Fundamental Research Funds for the Central Universities (grant numbers 3207022201A3, 4060692201/006) and High-level personnel project of Jiangsu Province. The authors thank the Center for Fundamental and Interdisciplinary Sciences of Southeast University for the support in MR measurement. The authors acknowledge the computational resources from the Big Data Center of Southeast University. The authors thank the computational resources from the Big Data Computing Center and the Center for Fundamental and Interdisciplinary Sciences of Southeast University.

**Author contributions:** R.W., X. L., B.Z., and H.W. contributed equally to this work. T.C. conceived and planned the project and experiments. T.C. and H.W. grew the $Fe_3Ga$ single crystals, and T.C., together with R.W., prepared the samples. R.W., B.Z., T.C., and H.W. carried out the transport and magnetization measurements and analyzed the data. L.L., C.X., R.W., Z.W., and T.C. performed the steady high-magnetic-field measurements and transport studies. R.W., X.G., and T.C. conducted the SEM/EDS and crystallographic-orientation analyses. X.L., L.M., and J.W. contributed the band-structure calculations. T.C. and K.X. wrote the manuscript. All authors discussed the results and contributed to the final version of the manuscript.

**Competing Interests Statement：** The authors declare no competing interests.

**Correspondence and Requests for materials should be addressed to**: chentaishi@seu.edu.cn; kexia@seu.edu.cn, liang.ma@seu.edu.cn.

## Figure Legends

**Fig. 1| Basic characterizations and ultra-low temperature RT in D0₃-Fe₃Ga.**

**a,** Band structure of D0₃–Fe₃Ga highlighting the formation of Weyl points with the lattice of c=6.13 Å. Semi-transparent solid lines within the dashed boxes mark the location of the tilted Weyl-crossing band extending deep into the flat-band region. The upper red arrow points to the 3D plot of the nodal web obtained before SOC inclusion, while the lower red arrow indicates SOC-induced Weyl points. The bottom-right panel shows the calculated AHC with SOC included. **b,** Photograph of the as-grown Fe₃Ga single crystals. **c,** The TEM image taken along the [111] crystallographic direction. **d,** Magnetization curves at 5 K and 300 K with the magnetic field applied along the [010] direction. **e,** The temperature dependence of the saturated magnetization on the Fe₃Ga single crystal, and the inset shows the high-temperature MT curve. **f,** The RT curve for Fe₃Ga single crystal. The insets respectively show RT at dilution temperatures, and the extracted exponent $n$ from fitting the data to the equation $\rho(T) = \rho_0 + AT^n$.

**Fig. 2| Giant anomalous Hall conductivity in D0₃-Fe₃Ga.**

**a,** Hall resistivity at different temperatures. The inset shows the configuration for Hall measurements. **b,** The calculated Hall conductivity at different temperatures using the formula: $-\sigma_{yx} = \frac{\rho_{yx}}{\rho_{xx}^2 + \rho_{xy}^2}$. **c,** The steady high-magnetic-field Hall and MR curves at 1.8 K, corresponding to the left and right axes, respectively. **d** The Drude two-carrier formula fit on the Hall resistivity at 1.8 K.

**Fig. 3| The robust flat-MR in D0₃-Fe₃Ga driven by chiral anomaly.**

**a,** The steady high-magnetic-field MR curves at 1.8 K. The flat-MRs are shown in red solid lines. The inset shows the measurement configuration **b,** the 2 K PHE and PLMR, which correspond to the left and right axes, respectively, and share the same coordinate scale; and the 300 K scenarios refer to the right axis. **c** and **d,** respectively, show the MR under $\widehat{\mathbf{B}} \perp \widehat{\mathbf{I}}$ and $\widehat{\mathbf{B}} \parallel \widehat{\mathbf{I}}$ at 2 K, 100 K, and 300 K.

**Fig. 4| Competition between chiral anomaly and anisotropic magnetoresistance.**

**a,** PLMR after subtracting the background at each temperature under a magnetic field of 9 T. To improve clarity, the curves are vertically offset with equal spacing, using the 2 K curve as the baseline. **b,** The amplitude ratio and phase difference between the PHE and PLMR . The inset shows the measurement configuration.

# Figures

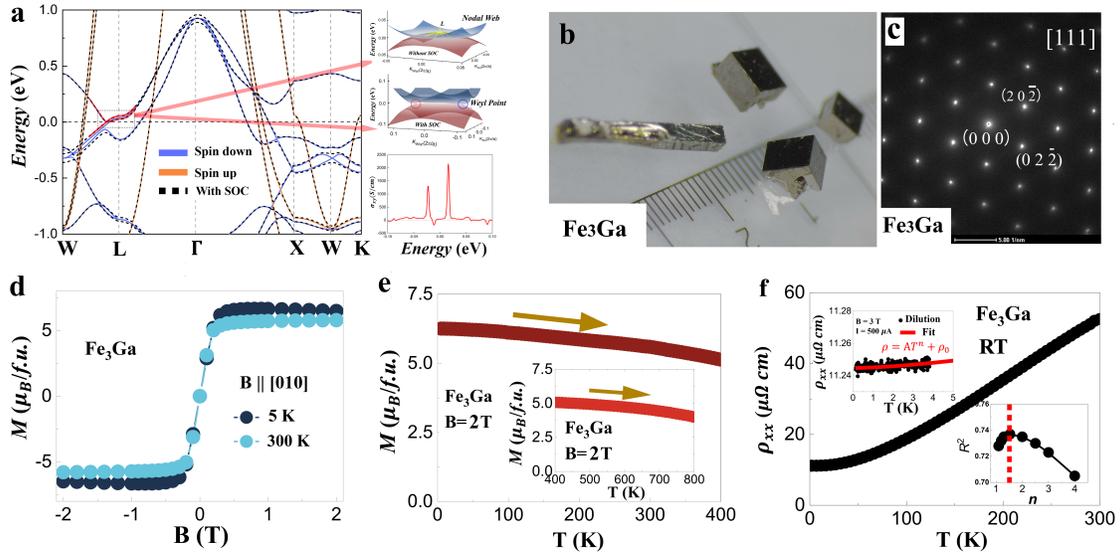

*Figure 1*

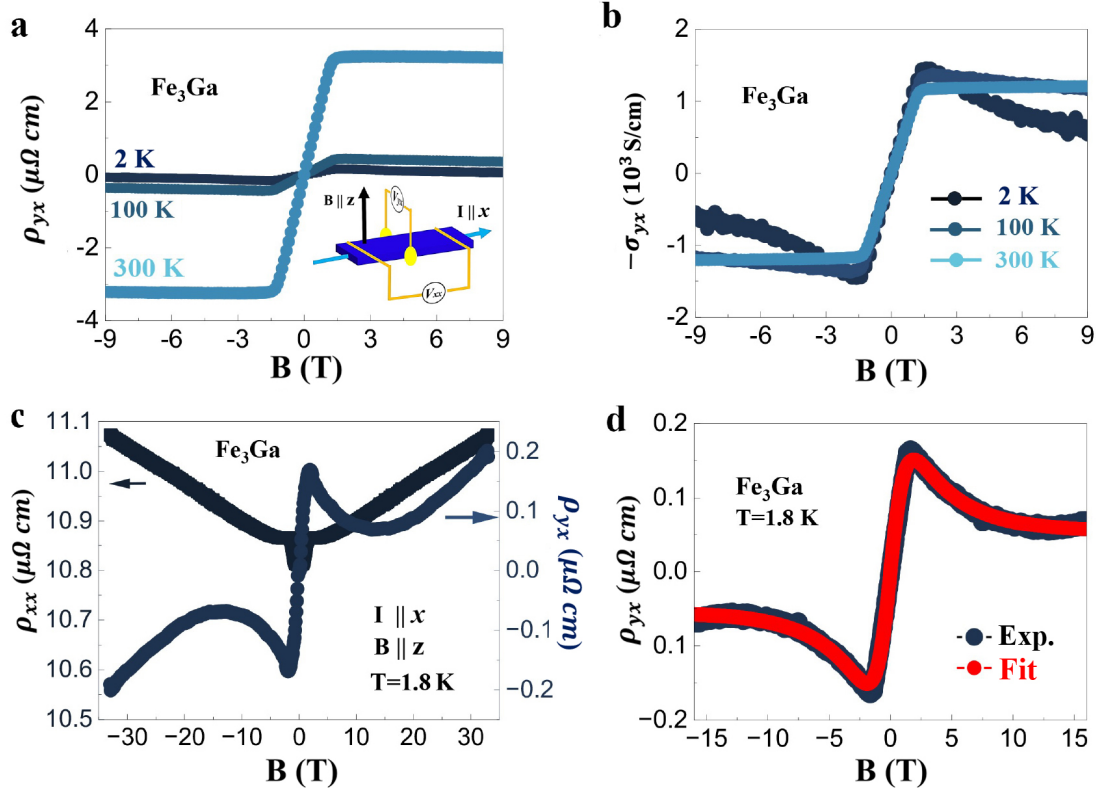

*Figure 2*

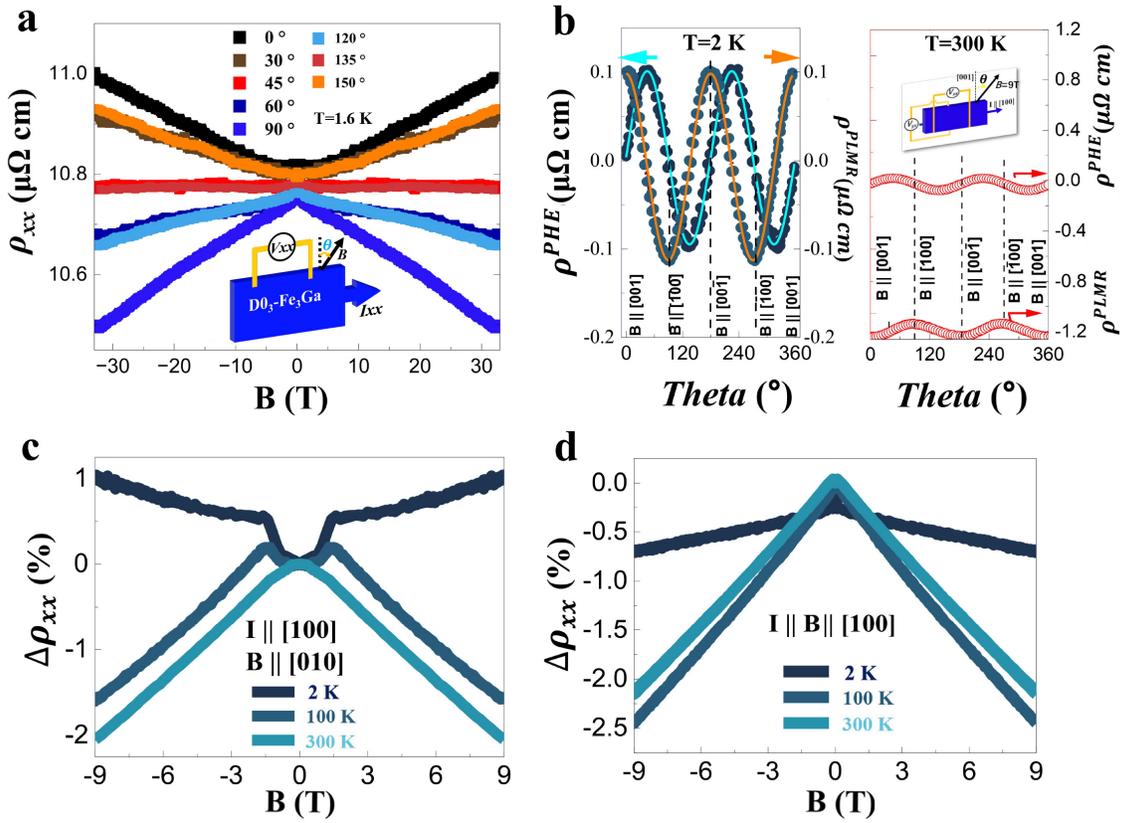

*Figure 3*

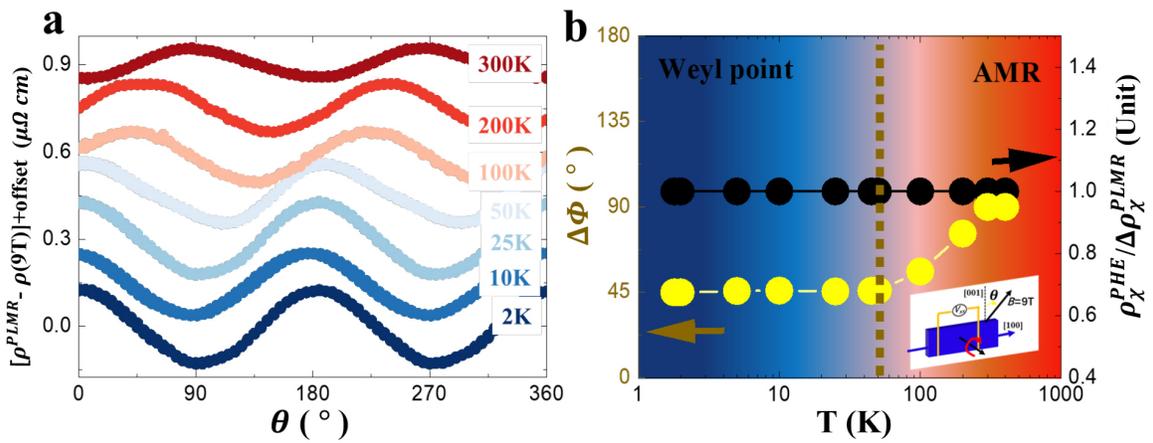

*Figure 4*

# Supplementary Information for: "Robust Flat-Magnetoresistivity in D0₃-Fe₃Ga Driven by Chiral Anomaly"

Ruoqi Wang[1]*, Xinyang Li[1]*, Bo Zhao[2,3]*, Haofu Wen[1]*, Xin Gu[1], Shijun Yuan[1], Langsheng Ling[4], Chuanying Xi[4], Ze Wang[4], Kunquan Hong[1], Liang Ma[1,5]✉, Ke Xia[1]✉, Taishi Chen[1]✉, Jinlan Wang[1,5]

1. *Key Laboratory of Quantum Materials and Devices of Ministry of Education, School of Physics, Southeast University, Nanjing 21189, China*

2. *Shanghai Advanced Research Institute, Chinese Academy of Sciences, Shanghai 201204, China*

3. *Henan Key Laboratory of Imaging and Intelligent Processing, Information Engineering University, Zhengzhou 450001, China*

4. *Anhui Key Laboratory of Low-Energy Quantum Materials and Devices, High Magnetic Field Laboratory, HFIPS, Chinese Academy of Sciences, Hefei, Anhui 230031, China*

5. *Suzhou Laboratory, Suzhou 215004, China*

## S1. The Scaling plot between Intrinsic AHC and Longitudinal Conductivity

In frame of single-partical, the intrinsic AHC at a finite temperature is expressed as: $\sigma_{xy}^{int.} = -\frac{e^2}{h}\sum_{n,k}\int\frac{dk}{2\pi^3}f(\varepsilon_n(k))\Omega_{n,z}(k)$ , where $f(\varepsilon_n(k))$ and $\Omega_{n,z}(k)$ refer to the Fermi-Dirac distribution function and Berry curvature[1, 2], respectively. The $\Omega_{n,z}(k)$ is given by $\Omega_{n,z}(k) = i\sum_{n\neq n'}\frac{<n'|v|n><n|v|n'>}{(\varepsilon_{n'}-\varepsilon_n)^2}$. Therefore, the $\sigma_{xy}^{int.}$ at finite temperature depends on the $\Omega_{n,z}(k)$. When a nondegenerate band-crossing occurs, the Berry curvature increases significantly, leading to a AHC in the order of $10^3$ S/cm, even in antiferromagnetic compounds[1, 2, 3, 4, 5, 6]. In previous work, the DFT calculation on D0₃-Fe₃Ga shows the intrinsic AHC peaks ~1200 S/cm at ~-10 meV below $\boldsymbol{E_F}$[7].

In ferromagnetic metals, the Hall resistivity $\rho_{xy}$ can be expressed as: $\rho_{xy} = R_0B + R_sM_s$, where $R_0, R_s$ , $M_s$ and denote the ordinary Hall coefficient, anomalous Hall coefficient and saturated magnetization[1], respectively. The intrinsic AHE resistivity, $\rho_{xy}^{int.} = R_sM_s$ , is quadratically proportional to the resistivity $\rho_{xx}$, namely $\rho_{xy}^{int.} \propto \rho_{xx}^2$. Using the tensor relations of resistivity, $\sigma_{xy}^{int.} = \frac{-\rho_{xy}}{\rho_{xy}^2 + \rho_{xx}^2}$, $|\sigma_{xy}^{int.}|$ behaves as a constant, independent of $\sigma_{xx}$, resulting in the scaling relation: $|\sigma_{xy}^{int.}| \propto |\sigma_{xx}^0|^{[1,2]}$.

Surprisingly, the experimentally measured AHC in our CVT- grown D0$_3$-Fe$_3$Ga is 200 S/cm higher than the peak value yielded by the DFT calculations in *Ref[7]*. Based on TEM measurements of the D0$_3$–Fe$_3$Ga single crystals grown via our chemical vapor transport (CVT) method, we determine a lattice constant $c = 6.13$ Å. Our calculations show the maximum intrinsic AHC to occur 18 meV above the calculated calculated Fermi level E$_F$= 0 eV, reaching an exceptionally large value of 2200 S/cm, the highest intrinsic AHC reported thus far from theoretical calculations. Combining our band-structure–based AHC calculations, the observed chiral-anomaly–induced negative MR, and the experimentally measured giant AHC, we infer that the Fermi energy in the CVT-grown D0$_3$–Fe$_3$Ga single crystals is at E$_F$ = 11 meV, lying almost exactly at the Weyl points (see Fig. S13).

Band-structure analysis reveals three key differences relative to earlier work[7]: (1) our lattice constant is 5.5% larger than that reported previously; (2) near the Fermi level, Weyl points arising from the crossing of flat bandsproduce strong Berry curvature; and (3) the Fermi surface simultaneously intersects the Weyl points and the narrow-gap flat bands.

The scaling plot for the $|\sigma_{xy}^{int.}|$ versus $|\sigma_{xx}^0|$ in Fe$_3$Ga is shown in Fig. S5. The giant $|\sigma_{xy}^{int.}|$ remains constant across all temperatures within the intrinsic regime, although the magnetic scattering dominates the MR at high temperatures. This strongly supports that the intrinsic AHC in Fe$_3$Ga originates from the 3D nodal flat-band rather than extrinsic mechanisms, such as the side-jump[1,2]. Furthermore, the Fe$_3$Ga samples grown by the CVT method exhibit an intrinsic AHC more than twice that of samples grown by the Czochralski method, despite their compositions being nearly identical. This can be attributed to the high quality of the CVT-grown Fe$_3$Ga samples. By comparing the TEM images, we clearly observe, we can clearly observe the satellite spots shown in the TEM image in *Ref[7]*. In contrast, the patterns in this work are sharp and clean. Defects that introduce additional charge carriers and shift the Fermi level have been studied in detail in the previous work[7].

## S2. The Comparison between AMR and Weyl Fermions Induced PLMR and PHE in D0$_3$-Fe$_3$Ga Single Crystal

To verify that the PLMR and PHE originate from the chiral anomaly, it is necessary to carefully account for the conventional AMR contribution, since Fe₃Ga is a typical 3*d* magnetic compound[8]. In conventional 3*d* transition metals, such as FeB thin films, the well-established expressions for planar AMR and PHE have been reported[9]:

$$\rho_{AMR}^{PLMR}(\theta) = \rho(0) + [\rho\left(\frac{\pi}{2}\right) - \rho(0)]cos^2(\frac{\pi}{2} - \theta) \qquad (S1)$$

$$\rho_{AMR}^{PHE}(\theta) = \frac{\left(\rho\left(\frac{\pi}{2}\right) - \rho(0)\right)}{2}\sin\left(2(\frac{\pi}{2} - \theta)\right) \qquad (S2)$$

where $(0)$ and $\rho\left(\frac{\pi}{2}\right)$ refer to the magnetoresistivity for $\widehat{\boldsymbol{B}} \perp \widehat{\boldsymbol{I}}$ and $\widehat{\boldsymbol{B}} \parallel \widehat{\boldsymbol{I}}$, respectively. Obviously, the formula (S1) shows the $\rho_{AMR}^{PLMR}$ maximum for $\widehat{\boldsymbol{B}} \perp \widehat{\boldsymbol{I}}$ and the minimum under $\widehat{\boldsymbol{B}} \parallel \widehat{\boldsymbol{I}}$, while the absolute value of $\rho_{AMR}^{PHE}$ is opposite. Figs. S6 and S7 compare the simulated $\rho_{AMR}^{PLMR}$ with the simulated Weyl-fermion-induced PLMR and the simulated $\rho_{AMR}^{PHE}$ with the Weyl fermions induced PHE, respectively. These results strongly support that the PHE and PLMR observed below 50 K originate from the chiral anomaly.

## S3. Squeezing-test Excluding Current-jetting Effect in D0₃-Fe₃Ga Single Crystal

The current-jetting effect refers to the inhomogeneous distribution of current density within a sample when an external magnetic field is applied parallel to the current direction. Owing to field-induced anisotropy in the electrical conductivity, carriers experience a suppressed transverse drift (perpendicular to B) compared to longitudinal drift, leading to the confinement of current into a narrow conducting channel inside the sample. This results in two key consequences: (1) as the applied field increases, the conductivity of the channel is relatively enhanced, producing an apparent negative MR that can mimic the signature of the chiral anomaly in topological semimetals[10]; and (2) the local conductivity becomes position-dependent across the sample. Given the resemblance of (1) to the intrinsic chiral-anomaly-induced negative MR, ruling out such artefacts is essential before interpreting experimental data.

An effective approach to eliminate this extrinsic contribution is the squeezing test. In this protocol, four terminal method measurements are performed with "point-contact" electrodes placed at different positions along the current path, with each contact diameter much smaller than the sample width *w* and length *l*. When $\widehat{\boldsymbol{B}} \parallel \widehat{\boldsymbol{I}}$, the MR is measured for multiple voltage-contact pairs:

if the MR curves from different positions show markedly different amplitudes, the presence of the current-jetting effect is inferred. Previous studies have shown that such artefacts are most pronounced in materials with ultrahigh carrier mobility, such as $Na_3Bi$[11].

In the present work, we applied the same squeezing test to D0$_3$- $Fe_3Ga$ single crystals (Fig. S8). Rectangular plate-like specimens with $l > w \gg t$ were prepared, and two sets of point-contact voltage leads were placed along the current direction, one near the left edge and the other at the center of the sample, as illustrated in the Fig. S8. Measurements were performed at T=2 K with $\hat{B} \parallel \hat{I}$. The results show that: (1) both voltage-lead pairs exhibit identical negative MR behavior, with normalized amplitudes in excellent agreement; and (2) the curves are consistent with those obtained earlier using current non-point-contact electrodes in the main text. Our observations unequivocally rule out current-jetting artefacts and confirm that the cryogenic negative MR arises from the intrinsic chiral anomaly.

### S4. From the Initial Eqs. 1–2 to the Final Eqs. 3–4: Detailed Derivation

As first established for Weyl semimetals by A. A. Burkov[12], the chiral-anomaly-induced PHE and PLMR can be expressed as follows:

$$\rho_{chiral}^{PHE} = -\Delta\rho_\chi \sin(\theta)\cos(\theta) \qquad (S3)$$

$$\rho_{chiral}^{PLMR} = \rho_0 - \Delta\rho_\chi \cos^2(\theta) \qquad (S4)$$

Here $\rho_0$ denotes the longitudinal resistivity as the longitudinal resistivity measured under zero magnetic field (B = 0), and $\Delta\rho_\chi$ is defined as $\Delta\rho_\chi = \rho_{xx}(0T) - \rho_{xx}(9T)|_{\hat{B}\parallel\hat{I}}$, and θ represents the angle between the applied magnetic field and the current direction within the Hall-current plane. Using standard trigonometric identities, Equations (1) and (2) can be rewritten in the following forms:

$$\rho_{chiral}^{PHE} = -\frac{1}{2}\Delta\rho_\chi \sin(2\theta) \qquad (S5)$$

$$\rho_{chiral}^{PLMR} = \rho_0 - \frac{1}{2}\Delta\rho_\chi(\cos2\theta + 1) \qquad (S6)$$

Expression (S6) is further transformed into:

$$\rho_{chiral}^{PLMR} = \rho_0 - \frac{1}{2}\Delta\rho_\chi \cos2\theta - \frac{1}{2}\Delta\rho_\chi \qquad (S7)$$

Expression (S7) is further transformed into:

$$\rho_{chiral}^{PLMR} = \rho_0 + \frac{1}{2}\Delta\rho_\chi \sin(2\theta - \frac{\pi}{2}) - \frac{1}{2}\Delta\rho_\chi \qquad (S8)$$

In our experiment, $\theta$ is defined as the angle between the magnetic field and the Hall direction, which differs from the definition used in $Ref^{12}$ by $\pi/2$. Therefore, in equations (S5) and (S8), $\theta$ should be replaced by $\frac{\pi}{2} - \theta$.

$$\rho_{chiral}^{PHE} = \Delta\rho_{\chi}\sin(2\theta) \qquad (S9)$$

$$\rho_{chiral}^{PLMR} = \rho_0 + \Delta\rho_{\chi}\sin\left(2\theta + \frac{\pi}{2}\right) - \frac{1}{2}\Delta\rho_{\chi} \qquad (S10)$$

Considering a fundamental observation from the D0₃–Fe₃Ga single-crystal experiment, Both the PHE and PLMR include simultaneous positive and negative MR components originating from the three-dimensional nodal flat band, whose magnitudes are equal. As a result, $\Delta\rho\chi$ is twice as large as the value obtained when only the negative MR component associated with the chiral anomaly is considered. Consequently, in Equations (S5) and (S8), $\Delta\rho_{\chi}$ should be multiplied by a factor of two. We finally arrive at the equations (3) and (4) given in the main text:

$$\rho_{modified}^{PHE}(\theta) = \Delta\rho_{\chi}\sin(2\theta) \qquad (S11)$$
$$\Delta\rho_{modified}^{PLMR}(\theta) = \Delta\rho_{\chi}\sin\left(2\theta + \frac{\pi}{2}\right) \qquad (S12)$$

It should be noted that Equations (3) and (4) in the main tex, although they successfully fit the PLMR and PHE data of D0₃–Fe₃Ga, are not universal, since their applicability requires that the magnitudes of the positive and negative MR components be approximately comparable. Nevertheless, to our knowledge, they constitute the first framework to simultaneously account for both effects within an ideal semimetal, rather than attributing PLMR and PHE solely to the chiral-anomaly-induced negative magnetoresistance commonly assumed in topological semimetals.

## S5. The Kadowaki–Woods Scaling Plot

The unified Kadowaki–Woods scaling plot, proposed by A. C. Jacko $et\ al.$ [13], provides a powerful Sframework for assessing the strength of electron–electron correlations in metals.

In the original work by Kadowaki and Woods[14], evaluating the correlation strength of different materials required simultaneous measurement of the low-temperature resistivity coefficient $A$ (from $\rho(T) = \rho_0 + AT^2$ ) and the Sommerfeld coefficient $\gamma$ from specific-heat data (from $C(T) = \gamma\ T$), thereby enabling the ratio $A/\gamma^2$ for comparative analysis.

However, the unified Kadowaki–Woods scaling proposed by Jacko et al. removes this constraint[13]. removes this constraint: in principle, once the band-structure information of a given system is known, measurement of either $A$ or $\gamma$ is sufficient to infer the other. This approach greatly streamlines the evaluation of correlation strength, especially when the limited sample amount precludes a reliable specific-heat measurement.

In our study of $D0_3$ - $Fe_3Ga$, repeated measurements left less than 1 mg of high-quality single-crystal material available, rendering accurate calorimetry impractical. Nevertheless, we obtained a low-temperature resistivity coefficient $A$=4.42$\times 10^{-4}$ $\mu\Omega$ cm/K² and and plotted this value on the unified Kadowaki–Woods scaling diagram (Fig. S9). The data show that $D0_3$- $Fe_3Ga$ lies more than an order of magnitude above that of typical transition metals such as Fe, Ni, Pd, Pt, and Re, indicating a markedly enhanced level of electronic correlations. This unusually strong correlation in the Fe-rich alloy $Fe_3Ga$ is particularly unexpected and is most likely attributable to the presence of flattened Weyl points..

Although the ultra-low-temperature resistivity exponent n = 1.53 cannot be directly derived from existing band-structure calculations or from current experimental data, the placement of our low-temperature resistivity coefficient A on the unified Kadowaki–Woods scaling plot suggests that the present sample already exhibits clear signatures of non-Fermi-liquid (NFL) behavior. Moreover, our DFT results, specifically the tilting of the Weyl cones and their relatively flat dispersion, provide a reasonable basis to infer that such flat bands may induce a certain degree of electronic correlation characterized by an effective interaction parameter *U*. A more detailed investigation into how the quantitative ratio U/W—between the correlation strength and the flat-band bandwidth—governs the emergence of NFL behavior will be pursued in future work.

## S6. The flat-MR happens at 45°and 135°

The flat-MRs occur at the expected 45° and 135°, but can also appear at other angles, such as 60° and 120°, due to the pronounced out-of-plane demagnetization effect[15, 16, 17]. In this study, we confirmed this demagnetization effect by performing angular-dependent MR measurements within the in-plane geometry, where demagnetization is minimized. Remarkably, the flat MR indeed occurs at 45° and 135°, in excellent agreement with theoretical expectations (see SI Fig. S10).

## S7. The DFT Calculation

In DFT calculations, $Fe_3Ga$ crystallizes in the D0$_3$-type structure with a lattice constant of 6.13 Å. The magnetic ground state is ferromagnetic, with a total magnetic moment of 6.27 $\mu_B$ per formula unit, and spin polarization oriented along the [100] direction of the primitive cell. The crystal structure employed in our DFT calculations is shown in Fig. S11.

Before including spin–orbit coupling (SOC), the calculated band structure at T = 0 K (Fig. 1a) displays a distinct band crossing along the **L–W** direction, forming a nodal web. Upon introducing SOC, the nodal web formed by the nodal lines is destroyed, and tilted Weyl points emerge. The corresponding conical dispersions appear within an extremely narrow energy window and extend into the adjacent flat-band region.

## S8. Wilson loop and Fermi arc confirming the Weyl point in D0$_3$–Fe$_3$Ga single crystal

Confirming Weyl points in the CVT-grown D0$_3$–Fe$_3$Ga single crystals requires particular caution. Previous work reported that the nodal web is destroyed upon inclusion of SOC, resulting in the opening of a full gap[7]. To verify this observation, we repeated the earlier calculations using the same lattice constant, c = 5.8 Å, and obtained identical results, including Weyl points located 0.9 eV below the Fermi level when SOC is present (Fig. S14).

In contrast, TEM measurements on our CVT-grown crystals yield a lattice constant of c = 6.13 Å. With this structural parameter, our DFT calculations reveal a clear band crossing along the **L–W** direction, persisting even after SOC is included. A 3D color-mapped rendering of the band dispersion confirms that this crossing is not a line-node intersection but rather an isolated point crossing.

To establish the topological nature of this point crossing, we employed the standard Wilson-loop method, which evaluates the accumulated Berry phase, also referred to as the winding-number method, of Wannier charge centers along closed surfaces enclosing each band-touching point. As shown in Fig. S12, the Wilson-loop evolution for the two enclosed crossings exhibits opposite winding directions, unambiguously identifying them as Weyl points, this opposite winding is the defining topological signature of Weyl-node chirality.

Another hallmark of Weyl points is the presence of surface Fermi arcs. Here, using calculated bulk- and surface-projected spectral weight, we resolve both the Weyl points and the connecting Fermi arcs, as shown in Fig. S12.

## S9. The Chiral Anomaly in $D0_3$- $Fe_3Ga$ Sample-2

As described in the Methods section, we screened more than 50 samples by measuring the residual-resistivity ratio (RRR) between 300 K and 2 K and selected the two highest-quality specimens: one with $RRR = 5.3$—the best sample used for all main-text figures and $RRR = 3.7$ (hereafter referred to as Sample 2). The remaining samples typically exhibited RRR values of approximately 2. Preliminary Hall-effect measurements showed that only these two highest-quality samples exhibited a clear Hall-slope transition (see Figs. S16b and S16c). Consequently, all key characterization experiments were focused on these two samples. Enhanced AHC and chiral-anomaly-induced negative MR were also observed, as shown in Figs. S16c and S16f. The AHC of Sample 2 is approximately 1300 S cm$^{-1}$, slightly smaller than that of the best sample reported in the main text. This close agreement strongly supports our conclusion that the unusual transport properties are intrinsic to high-quality D0$_3$–Fe$_3$Ga and not sample-specific.

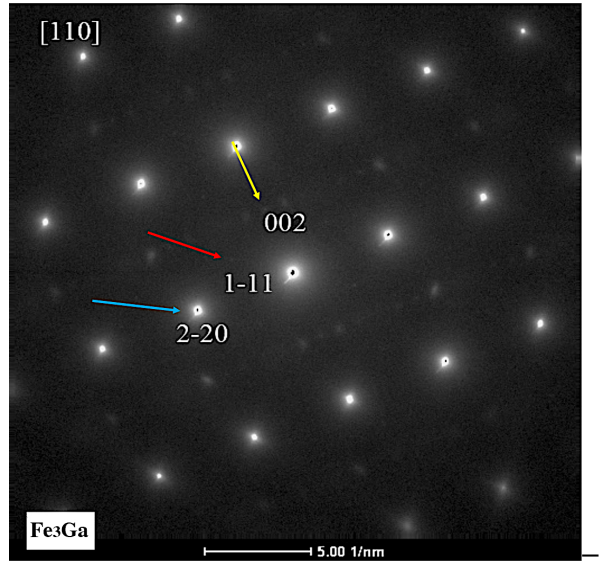

**Fig. S1| TEM results for Fe₃Ga single crystals.** TEM image collected from the crystallographic [110] direction. The yellow, red, and blue arrows refer to $(002)$, $(1\bar{1}1)$ and $(2\bar{2}0)$, respectively.

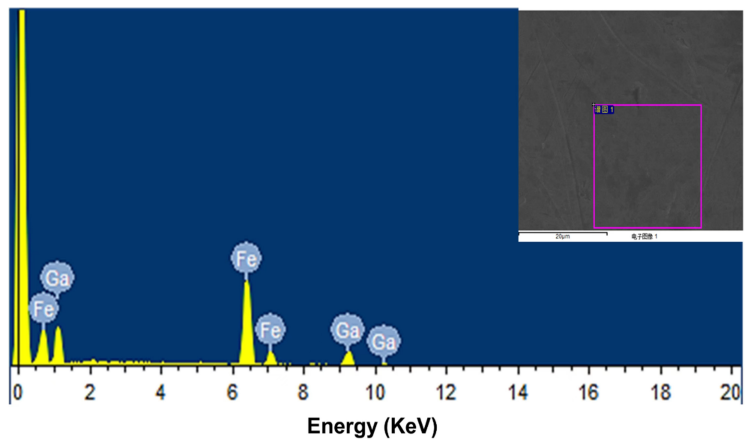

**Fig. S2| SEM/EDS analysis.** The average atomic ratio over 10 spots on the polished bulk yields the composition of iron and gallium 3.07:0.97.

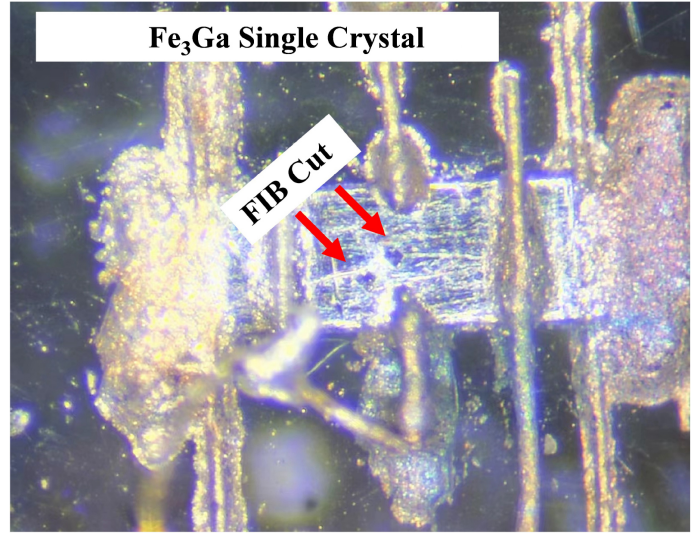

**Fig. S3| Fe₃Ga single crystal studied in the main text.** The two pits instructed by two arrows were made by a FIB-cut for TEM measurements, and these results are shown in Fig. 1c and Supplementary Fig. S1.

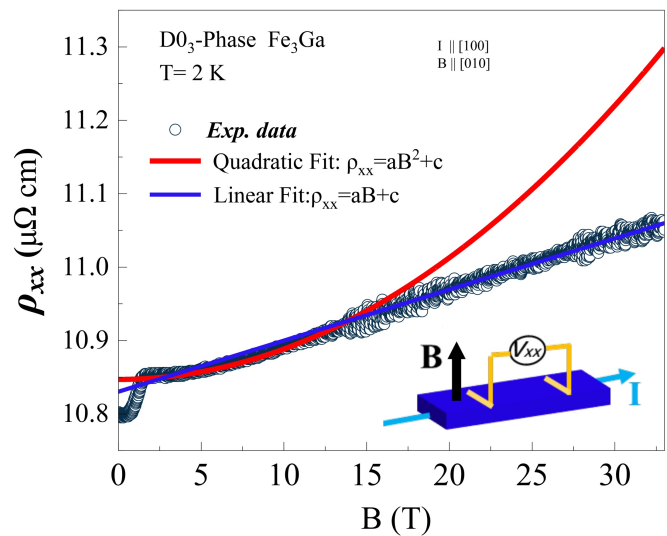

**Fig. S4| Increasing behavior of MR across different magnetic field ranges.** Low-field MR is fitted quadratically (red line), while high-field MR is fitted linearly (blue line).

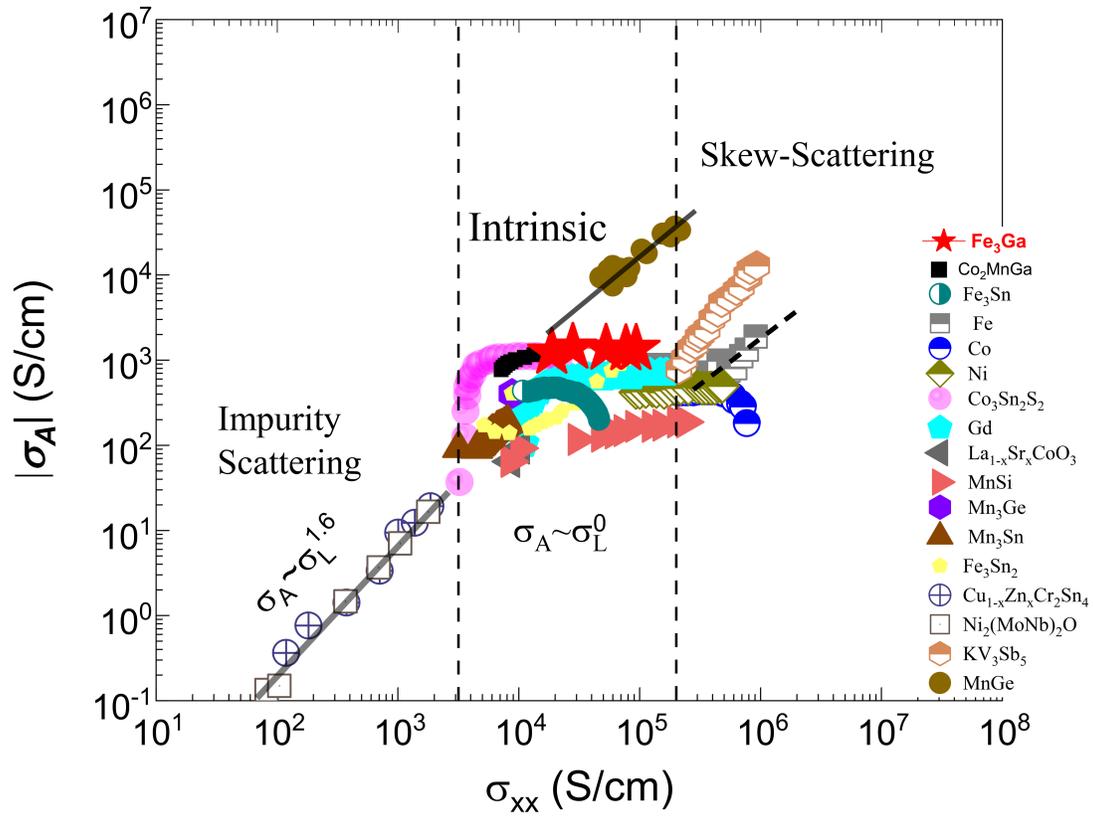

**Fig. S5| The scaling plot for the anomalous Hall conductivity and longitudinal conductivity.**
The pentagram symbol highlights our Fe₃Ga sample, located in the intrinsic regime.

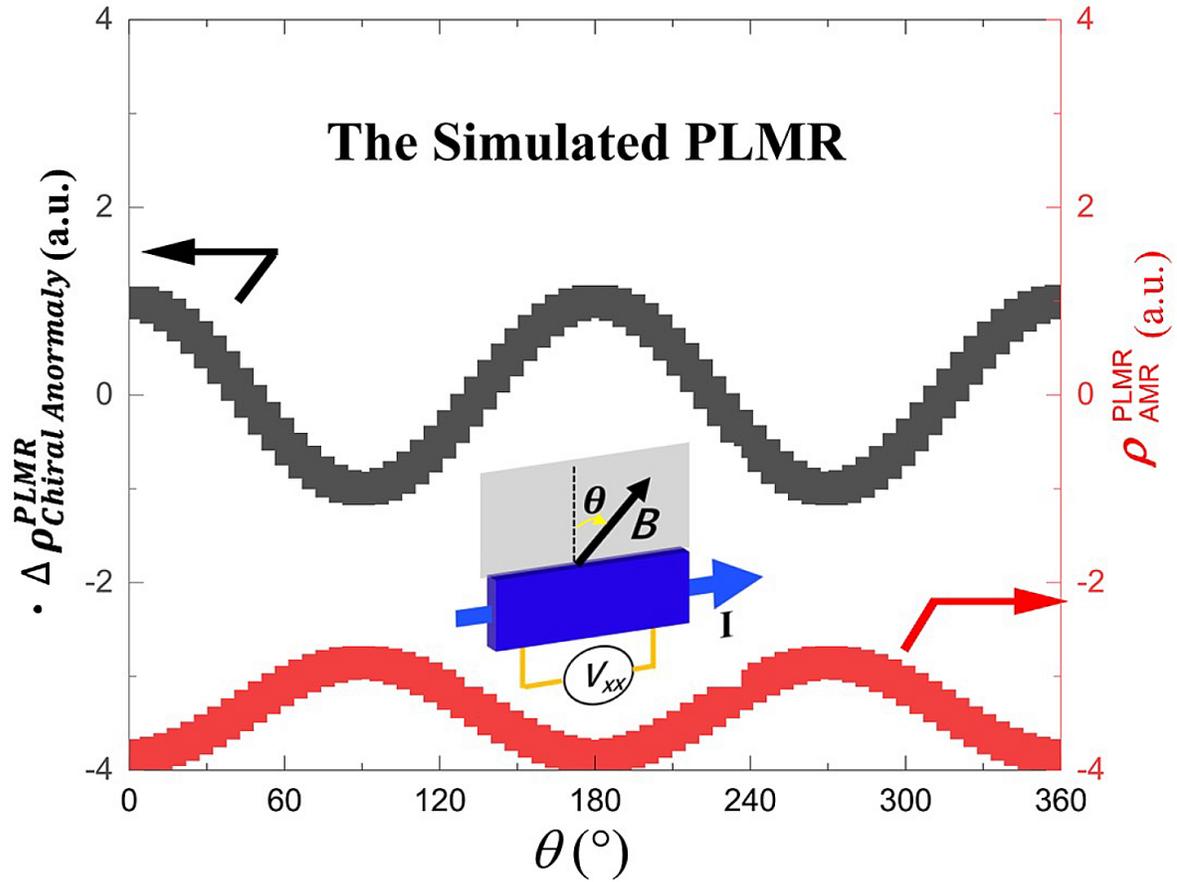

**Fig. S6| The simulated $\rho_{Chiral\ Anormaly}^{PLMR}$ and $\rho_{AMR}^{PLMR}$ induced by the chiral anormaly and AMR. The inset shows the configuration.** $\rho_{Chiral\ Anormaly}^{PLMR}$ is calculated using Eq. (S10), while $\rho_{AMR}^{PLMR}$ is derived from Eq. (S1). To enable a direct comparison of their angular dependence and overall lineshapes, both curves are normalized.

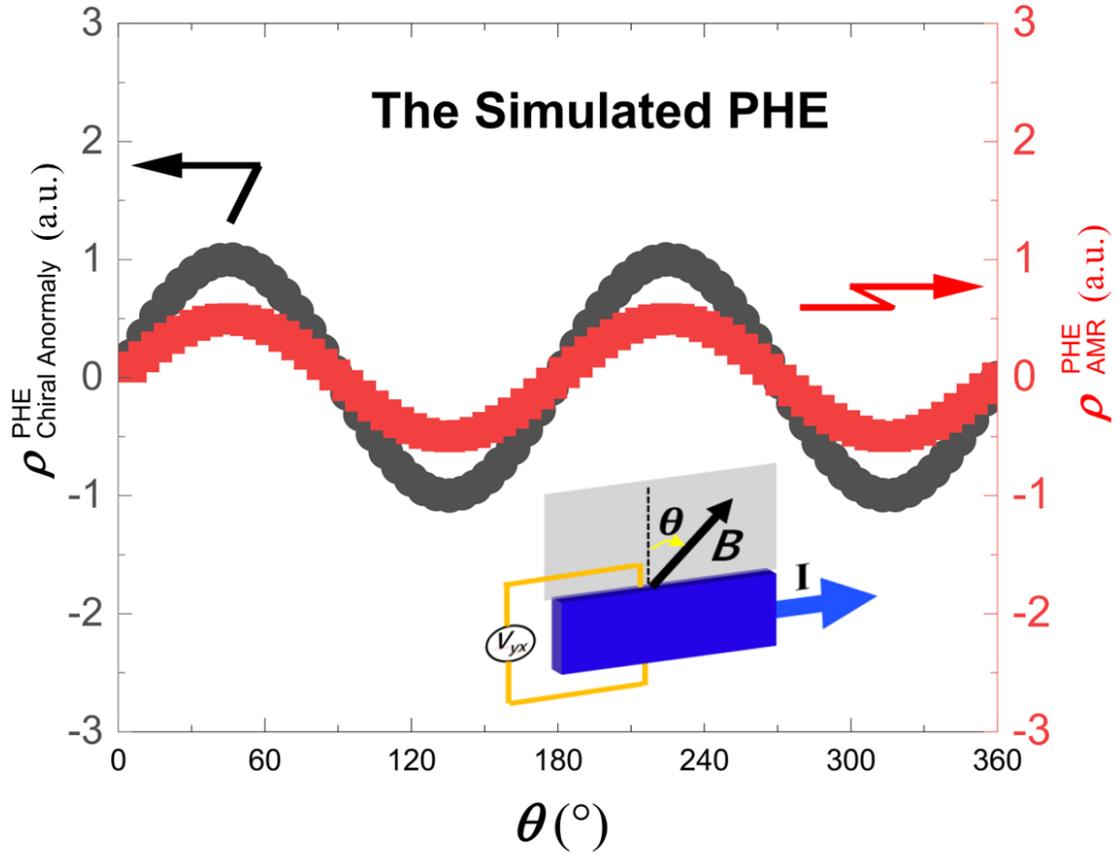

**Fig. S7|** The simulated $\rho_{Chiral\ Anormaly}^{PHE}$ and $\rho_{AMR}^{PHE}$ induced by the chiral anormaly and AMR. The inset shows the configuration. The inset shows the configuration. $\rho_{Chiral\ Anormaly}^{PHE}$ is calculated using Eq. (S9), while $\rho_{AMR}^{PHE}$ is derived from Eq. (S2). To enable a direct comparison of their angular dependence and overall lineshapes, both curves are normalized.

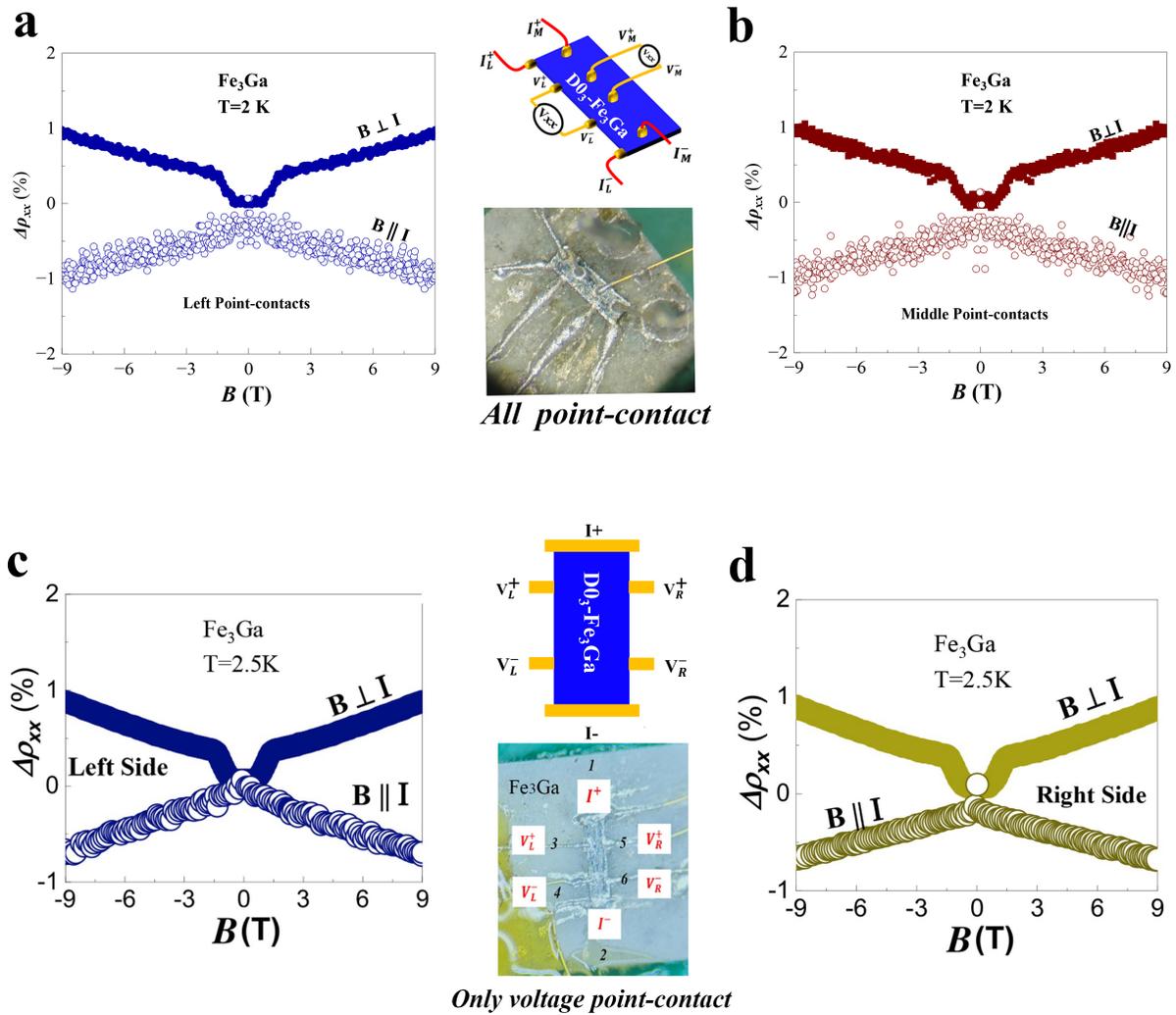

**Fig. S8| Squeezing test ruling out the current jetting effect.** a and b present the measurements obtained using the all-contact configuration, whereas c and d show results from the voltage-only contact configuration. The central panels illustrate the measurement geometries and the corresponding surface topographies employed in the squeezing test.

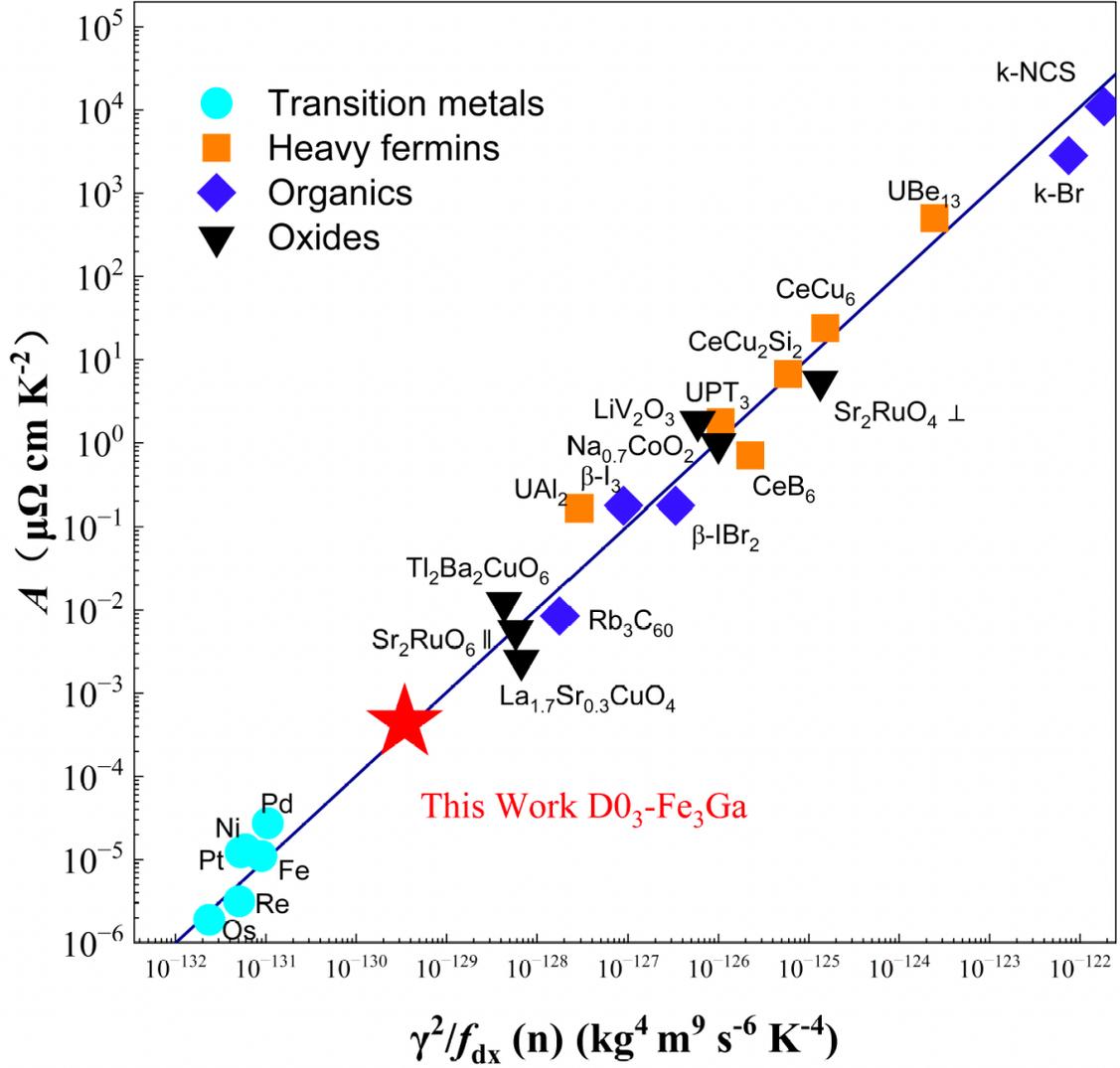

**Fig. S9| The electron-correlated compounds are summarized in the unified Kadowaki–Woods scaling plot.**

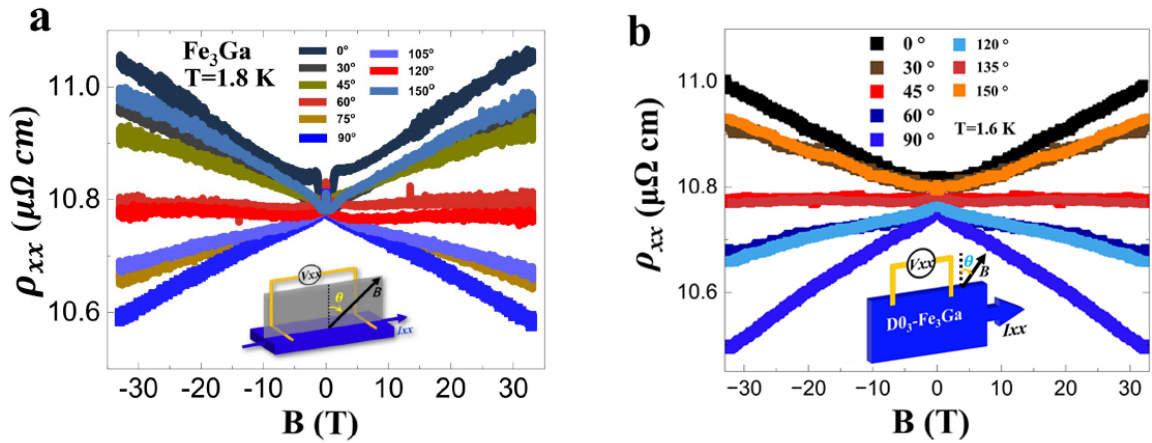

**Fig. S10 | Influence of the demagnetization effect on the angular-dependent MR of D0₃–Fe₃Ga.**

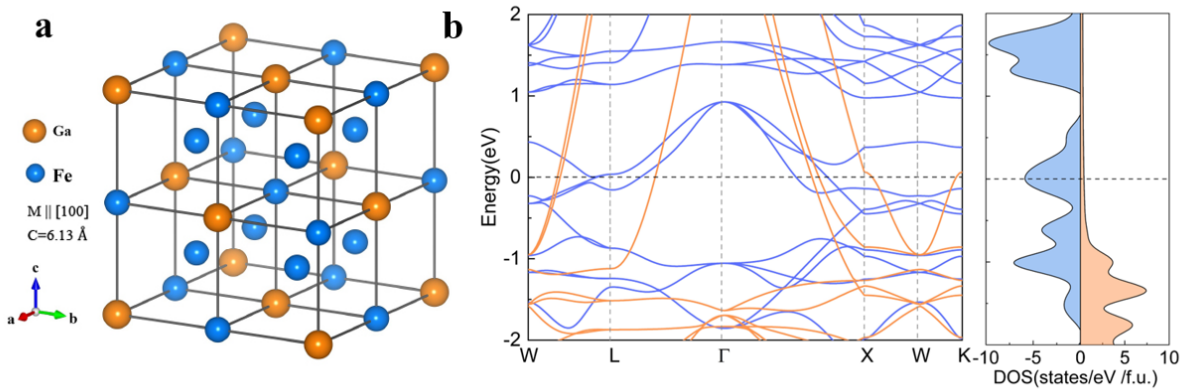

**Fig. S11| Crystal structure and Spin-polarized band structure and total density of states of D0₃-Fe₃Ga, using the lattice of c=6.13 Å.** Orange (blue) represents the up (down) spin channel.

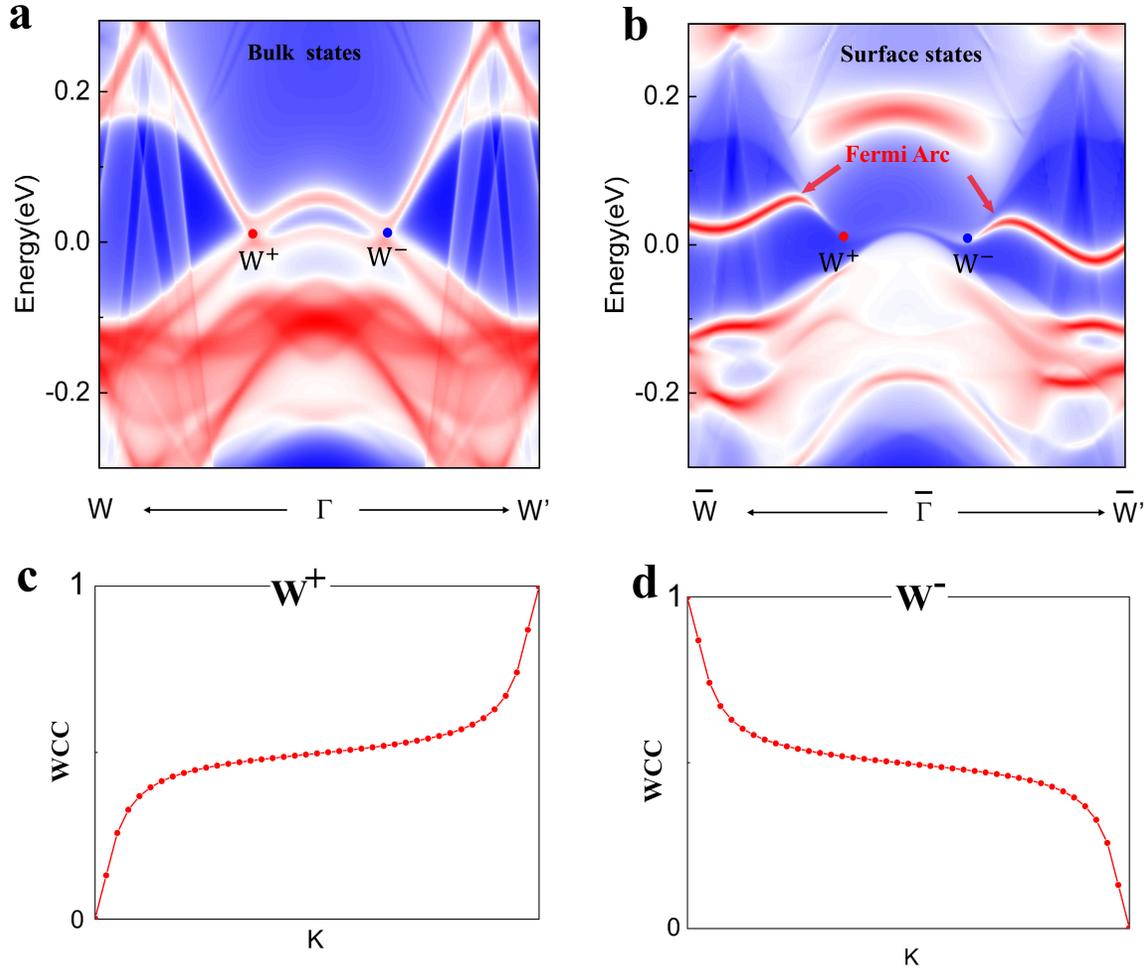

**Fig. S12 | Fermi arcs and Winding number analysis in D0₃–Fe₃Ga. a** and **b** compare the bulk and surface electronic structures near Weyl points. Red and blue denote the occupied and unoccupied states, respectively. **a** Projected bulk spectral function along the **W–Γ–W′** path. **b** Energy dispersion of the (001) surface state, highlighting the Fermi arcs connecting the surface projections of the Weyl points. **c** and **d** Wilson-loop evolution of the Wannier charge centers on closed surfaces enclosing the two SOC-induced band-touching points. The opposite winding directions of the loops unambiguously identify these crossings as Weyl points, as such winding is the defining topological signature of Weyl-node chirality.

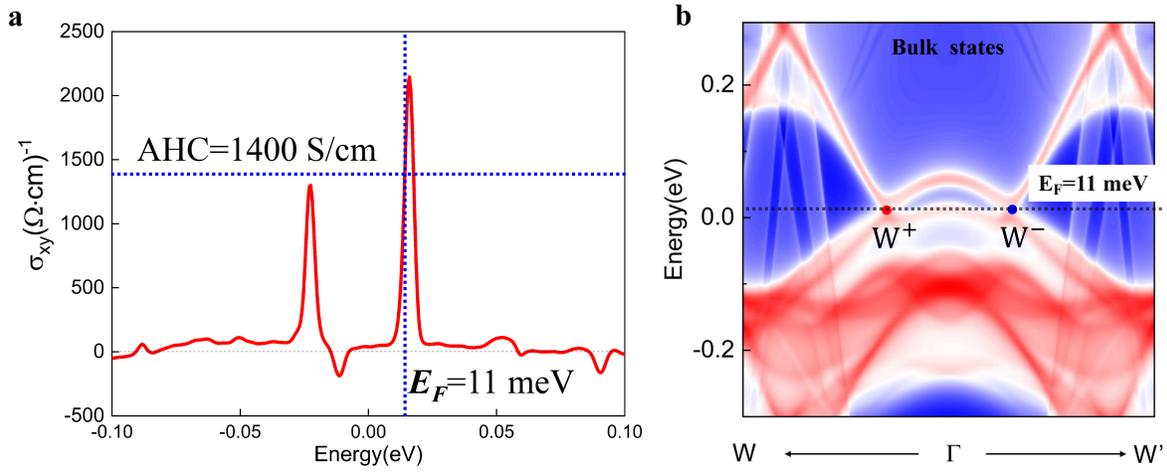

**Fig. S13 | Determination of the Fermi energy in D0₃–Fe₃Ga. a** Calculated anomalous Hall conductivity (AHC) as a function of Fermi energy, with SOC included. The blue dashed line denotes the position of the experimentally measured AHC, corresponding to a Fermi energy $E_F$ = 11 meV. **b** Band-structure view showing that $E_F$ = 11 meV intersects the Weyl points.

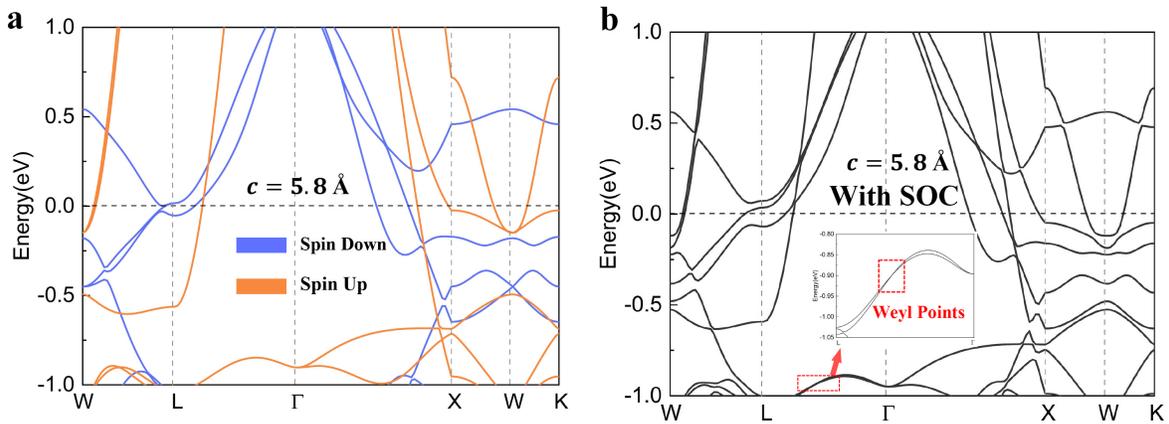

**Fig. S14| Weyl points in D0₃–Fe₃Ga with c = 5.8 Å, reproducing previous literature results.**

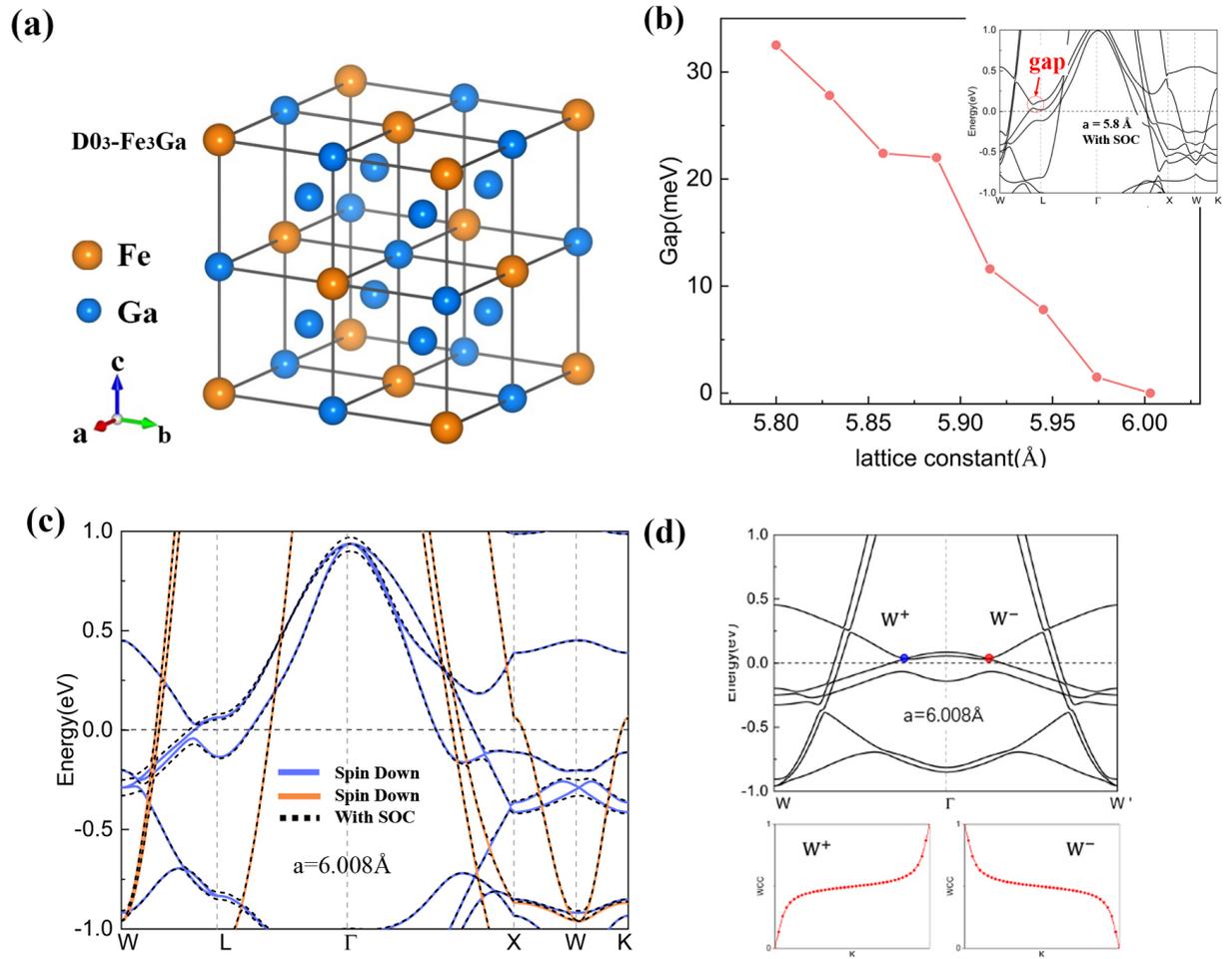

**Fig. S15 | Calculated evolution of the electronic band structure and emergence of Weyl points as the lattice constant of D0₃-Fe₃Ga expands.** By scanning the lattice constant to investigate the topological phase transition, we find that tensile strain effectively modulates the electronic band structure of the system. The band gap with increasing tensile strain and eventually closes at a critical strain of 3.5% (a=6.008 Å), thereby inducing tilt-Weyl points.

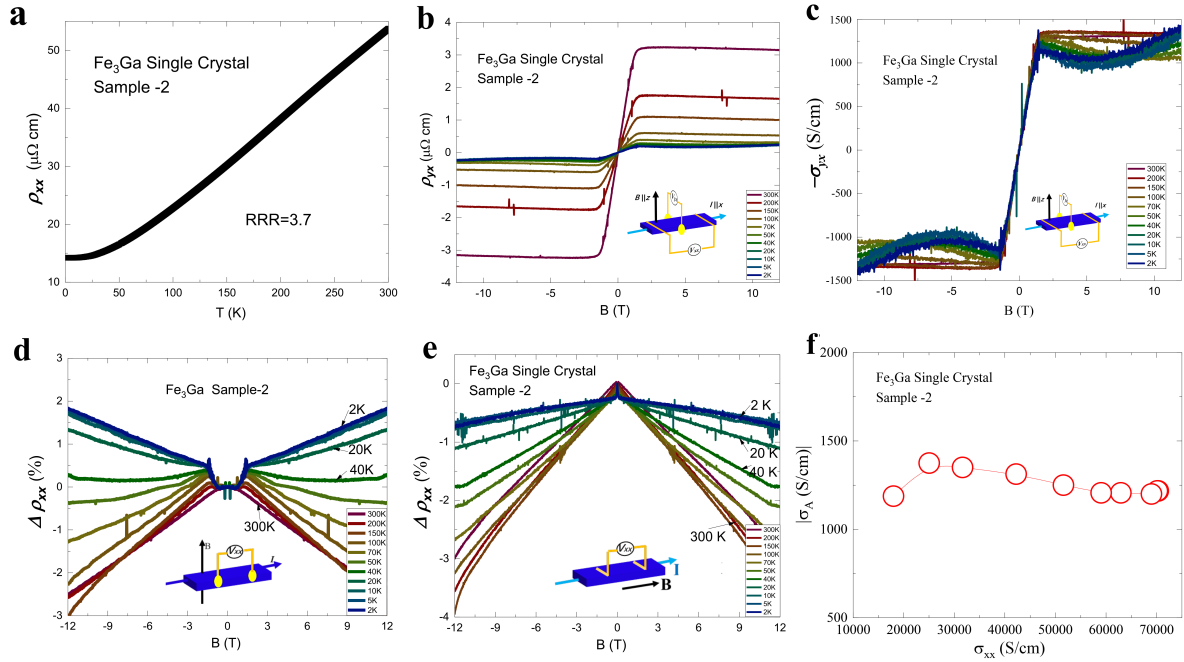

**Fig. S16| The information in the D0₃-Fe₃Ga Sample-2. a** The RT curve for Fe₃Ga single crystal sample-2. The RRR=$\rho(300K)/\rho(2K)$ equals 3.77, smaller than the 5 in the Fe₃Ga single crystal sample-1 discussed in the main text. **b** The anomalous Hall effect for Fe₃Ga single crystal sample-2 at different temperatures. **c** shows the corresponding anomalous Hall conductivity. (**d**) The delta magnetoresistivity under $\hat{B} \perp \hat{I}$ for Fe₃Ga single crystal sample-2 at different temperatures. **e** The delta magnetoresistivity under $\hat{B} \parallel \hat{I}$ at different temperatures in Fe₃Ga single crystal sample-2. **f** The scaling plot for the giant anomalous Hall conductivity versus Longitudinal conductivity in Fe₃Ga single crystal sample-2.

**Table: Systematic quantitative comparison with benchmark systems**

| Materials | Nodal type | Field range of negative MR (T) | Field of negative MR weakens markedly (Bw) | Anomalous Hall Conductivity (S/cm) | Temperature window of chiral anomaly (K) | Curie temperature (K) | References |
|---|---|---|---|---|---|---|---|
| **Na$_3$Bi** | Dirac | 0.3~2T (at 4.5K) | 2T | - | <100K | - | [11] |
| **TaAs** | Weyl | 1~5T (at 1.8K) | 5 | - | <75K | - | [18] |
| **Co$_3$Sn$_2$S$_2$** | Weyl | >14T (at 2K) | Not reported | 1200 | - | 175K | [19] |
| **Cd$_3$As$_2$** | Dirac | 1~5T (at 2K) | 5 | - | <300K | - | [20] |
| **GdPtBi** | Weyl | 2~5T (at 9K) | 5 | 60 | <75K | - | [21] |
| **Co$_2$MnGa** | Nodal line/Weyl | 0~12T (at 9K) | >16T | 1250~2000 | Not reported | 720K | [15, 22] |
| **Fe$_3$Ga** | **3D Nodal Flat band** | **>33T (<50K)** | **0.5 (2K 9T)** | **1400 (2K 9T)** | **<300K** | **>800K** | **Our work** |